\documentclass[
]{ceurart}

\usepackage{listings}
\usepackage{amsmath}
\begin{document}

\copyrightyear{2026}
\copyrightclause{Copyright for this paper by its authors.
  Use permitted under Creative Commons License Attribution 4.0
  International (CC BY 4.0).}

\conference{CLEF 2026 Working Notes,
  September 21 -- 24, 2026,  Jena, Germany}

\title{IROH: Insightful Ranking Of Humor using Multi-Stage Hybrid Retrieval with Rationale-Distilled LLM Judges for JOKER 2026 Track Task 1 English}

\title[mode=sub]{Notebook for the JOKER Lab at CLEF 2026}

\author[1]{Ana-Maria Luisa Mocanu}[%
orcid=0009-0005-2646-4086,
email=ana_maria.mogoase@upb.ro]
\fnmark[1]
\address[1]{National University of Science and Technology POLITEHNICA Bucharest, Splaiul Independen\c{t}ei 313, Bucure\c{s}ti 060042,
Romania}
\address[2]{Academy of Romanian Scientists, Ilfov 3, Bucharest, 050044, Romania}

\author[1]{Sebastian Mocanu}[%
orcid=0009-0007-0313-4724,
email=sebastian.mocanu@upb.ro
]
\fnmark[1]

\author[1,2]{Ciprian-Octavian Truic{\u{a}}}[%
orcid=0000-0001-7292-4462,
email=ciprian.truica@upb.ro,
url=https://sites.google.com/view/ciprian-octavian-truica,
]

\author[1]{Elena-Simona Apostol}[%
orcid=0000-0001-6397-4951,
email=elena.apostol@upb.ro,
url=https://sites.google.com/view/elena-simona-apostol,
]
\cormark[1]

\cortext[1]{Corresponding author.}
\fntext[1]{These authors contributed equally.}

\begin{abstract}
  Our team, VANGUARD, presents IROH (Insightful Ranking of Humor), a three-stage retrieval system for JOKER Task~1 English at CLEF~2026, achieving \textbf{first place} on the leaderboard with $0.6347$ MAP. Our pipeline combines hybrid sparse-dense retrieval, cross-encoder reranking, and a LoRA-adapted Large Language Model judge ensemble. We employ Gemma~4 to generate query-aware rationales under two prompt strategies, generic and typed, and produce up to four types of structured hard negatives for training data construction. Through an ablation across three cross-encoder architectures, four dense embedders, and eight judge configurations, our key findings are threefold: (1) the rationale-distilled judge is the primary driver of ranking quality, whereas appending rationales to the first-stage index contributes negligibly; (2) structured hard negatives degrade generalisation in nearly all configurations despite inflating local validation scores; and (3) across the components we ablate, the lighter, better-calibrated model is competitive with or stronger than its larger counterpart, with the generic-rationale Qwen2.5-7B judge ($0.6055$ MAP) outperforming every Gemma-4-31B configuration, and the advantage of generic over typed rationales is concentrated almost entirely in the smaller model.
\end{abstract}

\begin{keywords}
  humor retrieval \sep
  information retrieval \sep
  LLM judge \sep
  rationale distillation \sep
  cross-encoder reranking \sep
\end{keywords}

\maketitle

\section{Introduction}
Humor is a nuanced form of human language: it can be ambiguous, culturally loaded, ironic, or carried by tone alone. These properties make it difficult even for humans, and a particularly hard challenge for a literal computational ``brain''. Iroh once said, \textit{``It is important to draw wisdom from many different places''}~\cite{avatar2005}, a principle that guides our multi-stage design. Task~1~\cite{joker-1,joker-lncs} offers a natural language query that describes a humor topic, retrieving all the relevant jokes, puns, and wordplays from a balanced corpus of humorous and non-humorous texts.

We present IROH (Insightful Ranking of Humor)\footnote{Code available at \url{https://github.com/DS4AI-UPB/VANGUARD-CLEF2026-JOKER}.}, a three-stage retrieval system designed around the dual requirement of the JOKER task: the model must be aware of both semantic relevance and the specific linguistic typology that makes a text humorous. Our pipeline \textbf{ranks first place} on the leaderboard for Joker~2026 Track Task~1 English and contains:
\begin{itemize}
    \item Query-aware rationales and structured hard negatives generated with Gemma~4~\cite{gemma4_2026} for training-data construction, under two prompt strategies: a lightweight \textit{generic} pipeline producing three types of hard negatives, and a more structured \textit{typed} pipeline producing four hard negatives.
    \item A hybrid retrieval stage that combines BM25 with query expansion and dense BGE embeddings, fused via Reciprocal Rank Fusion (RRF), to cast a wide initial net. BM25 indexing is further enriched by appending generated rationales to documents, though we find this has little effect on retrieval performance (Subsection~\ref{ssec:discussion}).
    \item A finetuned cross-encoder that reranks the candidates for each query.
    \item A weighted ensemble of three LoRA-adapted judge instances, built from two base models: Qwen2.5-7B~\cite{hui2024qwen2} and Gemma~4-31B~\cite{gemma4_2026}, the latter finetuned separately on generic and typed rationales. This produces soft YES/NO scores that are combined via weighted voting to decide whether each candidate is a relevant humorous text.
    \item Across all three stages, scores are carried forward and fused via weighted linear interpolation of min-max normalised signals, with an optional confidence penalty at the final stage which, as reported in Subsection~\ref{ssec:post_processing_grid_search}, had no measurable effect on the final metrics.
\end{itemize}

While previous models successfully detect humor by capturing incongruity and structural features~\cite{suls1972two,weller2019humor,xie2021uncertainty}, the JOKER 2024 and 2025 tracks demonstrated that standard retrieval pipelines fail when both topical relevance and humor are required~\cite{ermakova2024overview,ermakova2025overview}. Prior leading approaches rely on post-hoc filtering~\cite{schuurman2024university} or zero-shot Large Language Model (LLM) classification~\cite{vachharajani2025pjmathematician}. In contrast, we propose a three-stage pipeline driven by a finetuned LLM judge.

To optimize our cross-encoder step, we build upon a multi-stage retrieval architecture~\cite{nogueira2019passage,zhuang2023rankt5} and the necessity of hard-negative mining~\cite{xiong2020approximate}. We then generate four structurally distinct humor-specific hard negatives inspired by LLM-driven augmentation approaches~\cite{li2024syneg}, tailored to combat the unique failure modes of humor retrieval systems. Instead of relying on zero-shot or few-shot LLM judges~\cite{rahmani2024llmjudge,niu2024judgerank}, we employ rationale distillation~\cite{hsieh2023distilling,tian2025beyond} to train a specialized humor-aware evaluator.

Our main contributions are:
\begin{itemize}
    \item A rationale-distillation approach under two prompt strategies (i.e., generic and typed), demonstrating that the simpler generic pipeline outperforms the more structured typed variant.
    \item An empirical study of structured hard-negative generation, revealing that augmentation inflates local validation scores while degrading official evaluation performance in nearly all configurations.
    \item An extensive ablation showing that, across the components we vary, model calibration and training-data composition matter at least as much as raw model capacity, with the lighter model matching or outperforming its larger counterpart.
    \item A pipeline configuration analysis revealing that candidate pool size governs recall depth, the CE/judge blend weight is a secondary knob, and the judge is the dominant top-rank signal.
\end{itemize}

The paper is organized as follows. Section~\ref{sec:Rw} reviews the related work. Section~\ref{sec:data} presents our data analysis. Section~\ref{sec:methodologies} describes the methodology and training procedures. Section~\ref{sec:results} reports the experimental results and ablation studies. Section~\ref{sec:conclusions} concludes with a summary and future directions.

\section{Related Work} \label{sec:Rw} This section addresses humor-aware information retrieval and focuses on three main categories: computational humor detection, multi-stage retrieval with neural reranking, and LLM-based knowledge distillation. 

Humor remains a fundamentally hard problem even for state-of-the-art LLMs. Recent work shows that even frontier models such as ChatGPT, Claude, and DeepSeek cap at roughly 51\% when identifying humorous punchlines in stand-up comedy transcripts, a ceiling that holds regardless of prompt engineering and underscores how brittle humor understanding remains~\cite{romanowski2025punchlines}. The multi-layered nature of humor partly explains this ceiling: incongruity resolution, tonal ambiguity, and cultural context interact in ways that resist decomposition into learnable surface features~\cite{suls1972two,xie2021uncertainty}. In retrieval settings, this challenge is further amplified, since a system must not only understand humor but rank it against a large corpus of plausible non-humorous distractors, a joint requirement that standard pipelines have been shown to handle poorly~\cite{ermakova2024overview}. Multi-stage retrieval architectures address recall and precision jointly by separating the problem into a broad first-stage retrieval and a more precise reranking step~\cite{nogueira2019passage,zhuang2023rankt5}. 

Hard-negative mining is essential for training effective rerankers, as it forces the model to distinguish genuinely relevant documents from superficially similar distractors~\cite{xiong2020approximate,meghwani2025hard}. LLM-generated synthetic negatives offer a scalable path to domain-specific augmentation~\cite{li2024syneg}, though their effectiveness relative to corpus-based mining remains dataset-dependent and does not always translate across evaluation distributions~\cite{sinha2025don}. In our setting, we observe that augmentation consistently inflates local validation scores while degrading official performance, which we attribute to a distribution shift.

Finally, rather than relying on zero-shot or few-shot LLM judges~\cite{rahmani2024llmjudge,niu2024judgerank}, which offer limited domain specificity, we employ rationale distillation~\cite{hsieh2023distilling,tian2025beyond} to train a specialized evaluator on humor-labeled query-document pairs enriched with linguistic rationales, whose calibration and interaction with the retrieval pipeline we analyse in Section~\ref{sec:results}, building on recent findings on LLM ranker-judge dynamics in IR evaluation~\cite{balog2025rankers}.

\section{Data Analysis} \label{sec:data}
The JOKER Task~1 corpus~\cite{joker-1} consists of short English texts with binary relevance judgments. We combine the 2025 and 2026 editions and deduplicate documents that recur across them, and after matching documents to the training qrels we construct a balanced dataset of jokes (label~1) and non-jokes (label~0) for model training.

Table~\ref{tab:data_stats} reports linguistic statistics by label. The strongest single correlate of the label is punctuation density ($r = 0.47$), which is more than twice as high in jokes ($0.424$ vs.\ $0.179$), reflecting their dialogue-heavy structure (e.g., \textit{``I've mailed the letter,'' Tom assented}). Closely related, jokes are far more likely to contain quotation marks ($56.2\%$ vs.\ $17.4\%$ with $r = 0.40$). These two signals are themselves strongly correlated with an $r = 0.65$, indicating they capture a shared dialogue-structure cue rather than independent features. Jokes are also markedly shorter than non-jokes (mean $10.5$ vs.\ $18.9$ words with $r = -0.31$) and carry more question marks on average ($0.126$ vs.\ $0.072$ per text). Overall, surface features are useful but insufficient. The remaining difficulty lies in the subtle, context-dependent wordplay that no single feature captures.

\begin{table}[h!]
\centering
\caption{Linguistic feature statistics by label.}
\label{tab:data_stats}
\begin{tabular}{lcccc}
\toprule
\textbf{Feature} & \multicolumn{2}{c}{\textbf{Joke (1)}} & \multicolumn{2}{c}{\textbf{Non-Joke (0)}} \\
\cmidrule(lr){2-3} \cmidrule(lr){4-5}
& Mean & Std & Mean & Std \\
\midrule
Word count          & 10.55 & 4.97  & 18.89 & 17.71 \\
Avg.\ word length   & 5.76  & 0.85  & 6.05  & 0.80  \\
Punctuation density & 0.424 & 0.303 & 0.179 & 0.118 \\
Has quotes (\%)     & 56.2  & ---   & 17.4  & ---   \\
Question marks      & 0.126 & 0.337 & 0.072 & 0.258 \\
\bottomrule
\end{tabular}
\end{table}

No single surface feature is sufficient: punctuation density is the strongest correlate of the label, followed by quote presence and word count (Table~\ref{tab:data_stats}).

\section{Methodologies} \label{sec:methodologies}
\begin{figure}
[htbp]
    \centering
    \includegraphics[width=1\linewidth]{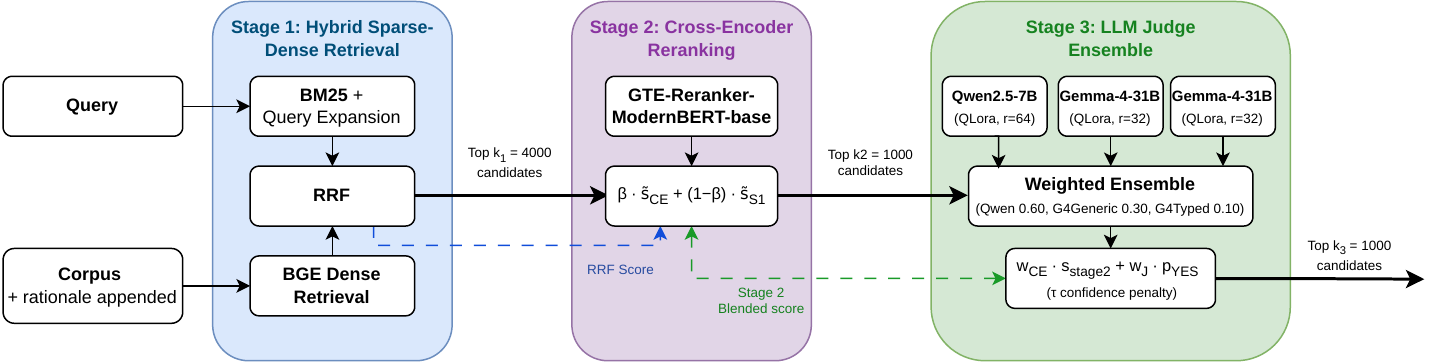} 
    \caption{Overview of the IROH three-stage retrieval pipeline.}
    \label{fig:pipeline}
\end{figure}

The IROH pipeline, illustrated in Figure~\ref{fig:pipeline}, processes a natural language query against a corpus of candidate documents enriched with generated rationales appended to their text. The first stage combines two parallel retrievers, BM25 with query expansion and BGE dense retrieval (\texttt{bge-base-en-v1.5}), whose ranked lists are fused via RRF as described in Equation~\eqref{eq:first-stage}, producing the top $k_1 = 4000$ candidates while carrying the RRF score forward. The second stage applies a finetuned GTE-Reranker-ModernBERT-Base cross-encoder to rescore these candidates, blending its output with the carried-forward RRF score as in Equation~\eqref{eq:second-stage}, yielding the top $k_2 = 1000$ candidates. In the third stage, three LoRA-adapted judge instances, one Qwen2.5-7B (QLoRA, $r=64$) and two Gemma-4-31B variants (QLoRA, $r=32$) finetuned on the generic and typed rationales, independently score each candidate; their soft YES probabilities are combined via a weighted ensemble (Qwen $0.60$, Gemma-generic $0.30$, Gemma-typed $0.10$), and the final ranking score is computed as described in Equation~\eqref{eq:last-stage}, applying a confidence penalty $\tau$ to suppress borderline negatives and producing the final ranked list of $k_3 = 1000$ documents. The complete prompts for rationale generation, hard-negative construction, and the LLM judge are listed in Appendix~\ref{app:prompts}.

\subsection{Rationale Generation}
We implemented two rationale generation pipelines, referred to throughout the experiments as \textit{generic} and \textit{typed}. Both use Gemma~4~\cite{gemma4_2026} via the Ollama framework, specifically the e4b variant, for efficient rationale generation. For each query-document pair, the model generates a one-sentence explanation of why the text is or is not a relevant joke, since LLMs struggle to learn humor from labels alone (e.g., for the text \textit{``\,`I bought myself fifty hamburgers, and I've only ten left,' said Tom with fortitude.''}, the model generated the rationale: \textit{``The text utilizes a pun by misapplying the word ``fortitude'' (meaning courage) to describe a simple quantitative statement, making the unexpected usage of the word itself the core source of the joke.''}). 

The \textit{generic} pipeline uses a ``General Wordplay'' placeholder as query context for all examples, keeping the prompt lightweight. The \textit{typed} pipeline retrieves the actual query text for each document from the training qrels, and uses a more structured prompt that explicitly enumerates seven humor mechanism types: homophonic puns~\cite{miller2017semeval}, homographic puns~\cite{miller2017semeval}, compound puns, Tom Swifties~\cite{lessard1992computational} that contain adverb-dialogue puns, double entendres~\cite{kiddon2011s}, malapropisms, and ironic twists. The seven types were selected to cover the major linguistic mechanisms present in the JOKER corpus, spanning phonological, semantic, structural, and lexical dimensions of wordplay without category overlap. For negative examples, both pipelines prompt the model to explain why the text lacks humor, mirroring the positive rationale structure, so the judge sees a symmetric supervision signal.

\subsection{Data Augmentation}\label{ssec:data_aug}
Aligned with the two rationale pipelines, we implement two augmentation strategies. The \textit{generic} pipeline generates three types of hard negatives per positive instance: literal rewrites (same topic, no humor), defused jokes (joke structure kept but punchline neutralized), and wrong-topic jokes (a genuine joke unrelated to the query). The \textit{typed} pipeline adds a fourth type, near-miss puns, which attempt wordplay related to the query but fail to produce a valid pun, making them harder negatives for the model to discriminate. Both pipelines inject the generated negatives into the training set (e.g. \textit{``\,`I bought myself fifty hamburgers, and I've only ten left,' said Tom with fortitude.''} $\rightarrow$ \textit{``Tom stated that he had purchased fifty hamburgers and that ten hamburgers remained.''}). The \textit{typed} pipeline additionally applies similarity-based quality filtering using character-level sequence similarity: literal rewrites are rejected if similarity to the original exceeds $0.85$, and defused jokes if it exceeds $0.95$. Exact-text deduplication is applied across the full augmented set after filtering.

\subsection{Hybrid Sparse-Dense Retrieval}
The first stage prioritises recall by combining complementary retrieval signals. We combine BM25 sparse retrieval with dense BGE embeddings using RRF~\cite{cormack2009reciprocal} as in Equation~\eqref{eq:first-stage}, where $r_i(q,d)$ is the rank of document $d$ under system $i$, $w_i$ is its assigned weight, and $k$ is a smoothing constant defaulted at $k=60$. BM25 is run once per query expansion variant and the element-wise maximum BM25 score across variants is retained before combining with dense, as it preserves the strongest lexical match signal without introducing fusion noise across variants. The top $k_1$ candidates advance to the second stage. BM25 indexing is further enhanced by appending the generated rationale to each document; as reported in Subsection~\ref{ssec:discussion}, this has a negligible effect on retrieval performance in isolation. The dense model selection is described in Subsection~\ref{ssec:exp_embedder}; we ultimately use \texttt{bge-base-en-v1.5} as detailed there.

\begin{equation}
    s_{\text{RRF}}(q, d) = \sum_{i} \frac{w_i}{k + r_i(q, d)}
\label{eq:first-stage}
\end{equation}

\subsection{Cross-Encoder Reranking}
We finetune three cross-encoder backbones to rerank the first-stage candidates: \texttt{ms-marco-MiniLM-L-6-v2} (MiniLM-L-6, 22M parameters) as a lightweight baseline, \texttt{BAAI/bge-reranker-base}~\cite{xiao2024cpack} (BGE-Reranker-Base, 278M parameters) as a higher-capacity alternative, and \texttt{Alibaba-NLP/gte-reranker-modernbert-base}~\cite{zhang2024mgte} (GTE-Reranker-ModernBERT-Base) as our primary candidate given its ModernBERT backbone. Two training configurations are evaluated: one using only the rationale-enhanced dataset with BM25 hard negatives and corpus-sampled negatives, and one additionally incorporating the structured hard negatives from Section~\ref{ssec:data_aug}. To prevent the cross-encoder from discarding relevant documents that scored highly in Stage~1, we blend its score with the Stage~1 RRF score using a weighted linear interpolation of min-max normalised scores as in Equation~\eqref{eq:second-stage}, where $\tilde{s}_{\text{CE}}$ and $\tilde{s}_{\text{S1}}$ are the min-max normalised cross-encoder and Stage~1 scores, and $\beta \in [0,1]$ controls the weight of the cross-encoder signal, selected via ablation as described in Subsection~\ref{ssec:piepline_ablation}. Only the top $k_2$ candidates advance to the final stage.

\begin{equation}
    s_{\text{stage2}}(q, d) = \beta \cdot \tilde{s}_{\text{CE}}(q, d)
    + (1 - \beta) \cdot \tilde{s}_{\text{S1}}(q, d)
\label{eq:second-stage}
\end{equation}

\subsection{LLM Judge}
The final stage applies a finetuned LLM judge ensemble to determine whether each candidate is a relevant humorous text. We finetune \texttt{Qwen2.5-7B-Instruct}~\cite{hui2024qwen2} and \texttt{gemma-4-31B-it}~\cite{gemma4_2026} using QLoRA, targeting all attention and feed-forward projection layers. Gemma is finetuned separately on the generic and typed rationale sets, yielding two Gemma judges; together with the Qwen judge this gives the three instances used in the ensemble (Subsection~\ref{ssec:ensemble}).

The judge's input is a system prompt for its role, the query, and the candidate text and its output is a binary YES/NO verdict. At inference time, we extract the logit probabilities of the YES and NO tokens and compute a soft score. Equation~\eqref{eq:last-stage} defines the final ranking score by linearly fusing the Stage~2 score with the judge's soft YES probability, with weights $w_{\text{CE}}$ and $w_{\text{J}}$. Documents whose $p_{\text{YES}}$ falls below a threshold $\tau$ receive an additional penalty factor $\lambda$, intended to suppress borderline negatives. This penalty had no measurable effect (Subsection~\ref{ssec:post_processing_grid_search}), and we retain it only for completeness. The top $k_3$ documents form the final submission.

\begin{equation}
    s_{\text{final}}(q, d) = w_{\text{CE}} \cdot s_{\text{stage2}}(q, d) + w_{\text{J}} \cdot p_{\text{YES}}(q, d)
\label{eq:last-stage}
\end{equation}

\subsection{Judge Ensemble}
\label{ssec:ensemble}

The three finetuned judges are combined via a weighted ensemble. Each judge independently scores every candidate, producing a soft YES probability $p_{\text{YES}}^{(j)}$ for judge $j$. The ensemble score is computed as a weighted sum as in Equation~\eqref{eq:ensemble}, where $w_j$ is the weight assigned to judge $j$ and $J$ is the set of judges. The final ranking score then fuses the ensemble output with the Stage~2 blended score as in Equation~\eqref{eq:last-stage}.

\begin{equation}
    p_{\text{YES}}(q, d) = \sum_{j \in J} w_j \cdot p_{\text{YES}}^{(j)}(q, d)
\label{eq:ensemble}
\end{equation}

The weight allocation is optimised via grid search as described in Subsection~\ref{ssec:post_processing_grid_search}, with the optimal configuration which assigns weight $0.60$ to Qwen and $0.30/0.10$ to generic and typed Gemma judges.

\section{Experiments} \label{sec:results}
We report official CodaBench scores as our primary metric throughout, computed with \texttt{pytrec\_eval} over the full corpus. A local held-out split of 20\% of queries is used for training-time model selection (early stopping); as shown in Subsection~\ref{ssec:exp_ce}, it diverges from official performance under augmentation.

\subsection{Training Details}
All experiments were run with PyTorch~2.x on Nvidia GPUs: an A100 80GB for judge finetuning and an RTX 4090 for the cross-encoders.

\paragraph{Cross-Encoder.} We finetune using the sentence-transformers~\cite{reimers2019sentence} library, with a fixed global seed of $42$ for reproducibility. For MiniLM-L6, we use a learning rate of $1 \times 10^{-5}$ and batch size $128$. For GTE-Reranker-ModernBERT-Base, memory constraints require a reduced batch size of $32$ and a learning rate of $2 \times 10^{-5}$. All cross-encoder models share cosine scheduling, $15\%$ warmup, weight decay $0.02$, and early stopping with patience $3$ on validation MAP, with a maximum of $50$ epochs. Training pairs are constructed from positive query-document pairs drawn from the rationale data, supplemented with random corpus negatives (2 per positive) and BM25 hard negatives (5 per query, mined from the top-80 retrieved candidates). Validation pairs are built independently with a separate fixed seed of $1337$, using 5 corpus negatives and 5 BM25 hard negatives per positive (mined from top-120), ensuring the validation signal is stable across runs. Early stopping is based on the mean average precision computed per query over the validation set.

\paragraph{Judge.} We finetune both judge models using QLoRA~\cite{dettmers2023qlora} with 4-bit NF4 quantization and double quantization enabled, using \texttt{bfloat16} compute dtype. LoRA adapters are applied with $r=64$ and $\alpha=128$ for Qwen2.5-7B-Instruct~\cite{hui2024qwen2}, and $r=32$ and $\alpha=64$ for Gemma~4-31B-it~\cite{gemma4_2026}, with a dropout of $0.05$, targeting the query, key, value, output, and feed-forward projection matrices. Training is performed using the TRL library's \texttt{SFTTrainer} with paged AdamW, cosine scheduling, $10\%$ warmup, and gradient checkpointing. For Qwen2.5-7B-Instruct~\cite{hui2024qwen2} we use learning rate $2 \times 10^{-4}$ and gradient accumulation over $8$ steps; for Gemma~4-31B-it~\cite{gemma4_2026} we use learning rate $5 \times 10^{-5}$ and gradient accumulation over $4$ steps. Both models use a maximum sequence length of $384$ tokens and a per-device batch size of $1$. Early stopping with patience $2$ on validation loss is applied, with a maximum ceiling of $30$ epochs.

\subsection{Cross-Encoder Comparison} \label{ssec:exp_ce}
To identify the best reranking model for Stage~2, we evaluate three cross-encoder architectures finetuned on our training data: \texttt{cross-encoder/ms-marco-MiniLM-L-6-v2}, \texttt{BAAI/bge-reranker-base}~\cite{xiao2024cpack}, and \texttt{Alibaba-NLP/gte-reranker-modernbert-base}~\cite{zhang2024mgte}. Each model is trained using the typed rationale-annotated training data only or with augmentation. The Stage~2 components are the same across all models, using rationale-expanded BM25 combined with dense retrieval.

Table~\ref{tab:ce_comparison} reports the results. Augmentation degrades CodaBench MAP for the two strongest backbones, consistent with an overfitting effect we observed on the local split, which improved under augmentation. The pattern suggests that the augmented examples shift the model's decision boundary towards the distribution of the local split rather than the broader retrieval task, resulting in inflated local scores. The best architecture for our task is GTE-Reranker-ModernBERT-Base trained without augmentation, outperforming all other configurations. MiniLM-L-6 is a competitive choice from a cost perspective, offering strong results relative to its size. Augmentation degrades performance for the two strongest backbones (MiniLM-L-6 and GTE-Reranker-ModernBERT-Base) with the sole exception being BGE-Reranker-Base, where both configurations score poorly, and the ordering reverses. For the architecture we adopt, the non-augmented, generic-rationale configuration is clearly the stronger choice.

Given the results, we select GTE-Reranker-ModernBERT-Base trained without augmentation as the Stage~2 cross-encoder for all subsequent experiments.

\begin{table}[h!]
\centering
\caption{Comparison of CodaBench MAP for cross-encoder models with and without data augmentation. Best values are \textbf{bolded}.}
\label{tab:ce_comparison}
\begin{tabular}{lcc}
\toprule
\textbf{Model} & \textbf{Augmentation} & \textbf{CodaBench MAP} \\
\midrule
MiniLM-L-6 & True & 0.1793 \\
 & False & 0.2215 \\
\midrule
BGE-Reranker-Base & True & 0.1229 \\
 & False & 0.0733 \\
\midrule
GTE-Reranker-ModernBERT-Base & True & 0.2597 \\
  & False & \textbf{0.2843} \\
\bottomrule
\end{tabular}
\end{table}

\subsection{Dense Embedder Comparison} \label{ssec:exp_embedder}
Following the same configurations described in Subsection~\ref{ssec:exp_ce} with the best cross-encoder, we compare four BGE-family models~\cite{xiao2024cpack,chen2024bge}: \texttt{bge-base-en-v1.5} (our baseline, 768-dim), \texttt{bge-large-en-v1.5} (1024-dim, same family), \texttt{bge-m3}~\cite{chen2024bge} (1024-dim, multi-granularity), and \texttt{bge-en-icl}~\cite{li2025making} (in-context learning embedder). Table~\ref{tab:embedder_comparison} reports CodaBench MAP for each model. \texttt{bge-en-icl}~\cite{li2025making} achieves the highest isolated result, but at substantially higher cost as a 7B-parameter LLM-based embedder. The performance gain is not attributable to the higher dimensionality but to the in-context learning mechanism, which encodes the humor retrieval objective directly into the query embedding. We select \texttt{bge-base-en-v1.5} for all subsequent experiments, as the full pipeline ablation in Subsection~\ref{ssec:piepline_ablation} shows \texttt{bge-base-en-v1.5} achieves higher MAP when paired with the judge ensemble.

\begin{table}[h!]
\centering
\caption{Dense embedder comparison. All runs use GTE-Reranker-ModernBERT-Base
(no augmentation) as the fixed Stage~2 cross-encoder.
Best values are \textbf{bolded}.}
\label{tab:embedder_comparison}
\begin{tabular}{lcc}
\toprule
\textbf{Model} & \textbf{Dimension} & \textbf{CodaBench MAP} \\
\midrule
bge-base-en-v1.5  & 768   & 0.2843 \\
bge-large-en-v1.5 & 1024  & 0.2823 \\
bge-m3            & 1024  & 0.2634 \\
bge-en-icl        & 4096  & \textbf{0.3290} \\
\bottomrule
\end{tabular}
\end{table}

\subsection{Stage-1 and Stage-2 Hyperparameter Search}
Having fixed the Stage~2 cross-encoder in Subsection~\ref{ssec:exp_ce}, we test the pipeline's sensitivity to the Stage~1 retrieval and Stage~1-Stage~2 fusion hyperparameters. We grid-search six parameters ($k_1$, the BM25 saturation $k_1^{\text{BM25}}$ and length normalization $b$, the RRF constant $k_{\text{RRF}}$, the BM25-to-dense weight $w_B$, and the number of query-expansion variants), yielding $324$ combinations in total. Given the CodaBench submission limits, we evaluated the top~15 configurations by local validation MAP; the best reached MAP $0.2905$. A follow-up One-Factor-At-A-Time sweep raised this to $0.2921$.

Both searches confirm that these parameters have a comparatively small effect ($0.2843 \to 0.2921$), an order of magnitude below what component selection moves: switching the embedder from \texttt{bge-base-en-v1.5} to \texttt{bge-en-icl} alone raised MAP from $0.2843$ to $0.3290$. The dominant levers are therefore the Stage~2 cross-encoder, the Stage~3 judge, and the dense embedder, not the first-stage retrieval and fusion hyperparameters.

\subsection{Judge Comparison} \label{ssec:judge_comparation}
To identify the best judges for Stage~3, we evaluate two models: Qwen2.5-7B-Instruct~\cite{hui2024qwen2} with LoRA $r=64$ and Gemma-4-31B-it~\cite{gemma4_2026} with LoRA $r=32$, each finetuned on four data variants: generic rationale only, generic rationale with augmentation, typed rationale only, and typed rationale with augmentation. For Stage~3, all runs use equal blend weights $w_{\text{CE}} = w_{\text{J}} = 0.50$, a confidence threshold $\tau = 0.40$, and $k_3 = 1000$ candidates. Table~\ref{tab:judge_comparison} presents the MAP for each configuration, using \texttt{CE\_GTE\_typed} as the fixed cross-encoder in all runs.

The results confirm that augmentation consistently degrades performance across both model families and that the generic rationale pipeline outperforms the typed variant for both judges despite the typed pipeline's richer prompt structure. The smaller Qwen2.5-7B judge trained on generic rationales achieves the best MAP of $0.6055$, ahead of every Gemma-4-31B configuration; this does not hold per-variant, however, as Gemma-4-31B is the stronger model on the typed rationales ($0.5740$ vs. $0.4933$). We revisit this in Subsection~\ref{ssec:discussion}.
 
\begin{table}[h!]
\centering
\caption{Judge model comparison. Best Codabench MAP in \textbf{bold}. When augmentation is used, it's the same type as the rationale.}
\label{tab:judge_comparison}
\begin{tabular}{llcc}
\toprule
\textbf{Model} & \textbf{Data} & \textbf{LoRA $r$} & \textbf{MAP} \\
\midrule
Qwen2.5-7B  & typed       & 64 & 0.4933 \\
Qwen2.5-7B  & typed + aug & 64 & 0.2987 \\
Qwen2.5-7B  & generic       & 64 & \textbf{0.6055} \\
Qwen2.5-7B  & generic + aug & 64 & 0.4994 \\
Gemma-4-31B & typed       & 32 & 0.5740 \\
Gemma-4-31B & typed + aug & 32 & 0.4538 \\
Gemma-4-31B & generic       & 32 & 0.5718 \\
Gemma-4-31B & generic + aug & 32 & 0.4267 \\
\bottomrule
\end{tabular}
\end{table}

\subsection{Pipeline Configuration Ablation} \label{ssec:piepline_ablation}
Having identified the best judge, Qwen2.5-7B trained on the generic rationale distillation process, with a MAP of $0.6055$ in Subsection~\ref{ssec:judge_comparation}, we investigate whether further gains are possible by varying the Stage~1 dense embedder, the candidate pool size ($k_1$, $k_2$), and the Stage~3 cross-encoder/judge blend weights, keeping all other components fixed. All runs use \texttt{CE\_GTE\_typed} as the cross-encoder and \texttt{Judge\_Qwen7B\_generic} as the judge.

Table~\ref{tab:pipeline_ablation} reports each configuration on the official CodaBench test queries. Run~1 is the baseline: \texttt{bge-base-en-v1.5}, equal blend weights $w_{\text{CE}}=w_{\text{J}}=0.50$, $k_2=500$ candidates, and a penalty threshold $\tau=0.40$. Run~2 replaces the embedder with \texttt{bge-en-icl}, keeping all other parameters identical. Run~3 takes Run~2 as a starting point and applies the OFAT-optimised Stage~1/Stage~2 parameters: BM25 $k_1^{\text{BM25}}=1.5$, RRF BM25 weight $0.8$, three query-expansion variants, Stage~2 CE blend $\beta=0.7$, and deduplication threshold $0.93$ (the pool size $k_1$/$k_2$ is unchanged from Run~2). Run~4 widens the candidate pool to $k_1=4000$, $k_2=1000$ ($\beta=0.85$, deduplication threshold $0.99$, $\text{QE}=5$). Runs~5-7 keep this wider pool and sweep the blend: Run~5 uses \texttt{bge-en-icl} at $0.35/0.65$. Runs~6-7 use \texttt{bge-base-en-v1.5} at $0.35/0.65$ and $0.45/0.55$ respectively ($\tau=0.30$). Run~8 pushes the blend further to $0.30/0.70$, widens Stage~2 to $k_2=1500$, and raises the penalty threshold to $\tau=0.35$.

\begin{table}[h!]
\centering
\caption{Pipeline configuration ablation. All runs use \texttt{CE\_GTE\_typed} and
\texttt{Judge\_Qwen7B\_generic}. Best values per metric are \textbf{bolded}.}
\label{tab:pipeline_ablation}
\setlength{\tabcolsep}{4pt}
\begin{tabular}{lllccccccc}
\toprule
\textbf{Run} & \textbf{Embedder} & $w_{\text{CE}}$/$w_{\text{J}}$ & $k_1$ & $k_2$ & \textbf{MAP} & \textbf{NDCG@10} & \textbf{NDCG@100} & \textbf{RR} & \textbf{rel\_ret} \\
\midrule
1 & bge-base   & 0.50/0.50 & 3000 & 500  & 0.6055 & \textbf{0.6836} & 0.7276 & 0.7990 & 3564 \\
2 & bge-en-icl & 0.50/0.50 & 3000 & 500  & 0.6009 & 0.6760 & 0.7417 & 0.8057 & 4029 \\
3 & bge-en-icl & 0.50/0.50 & 3000 & 500  & 0.5886 & 0.6538 & 0.6694 & 0.7824 & 3541 \\
4 & bge-en-icl & 0.50/0.50 & 4000 & 1000 & 0.5984 & 0.6749 & 0.7400 & 0.8075 & 4040 \\
5 & bge-en-icl & 0.35/0.65 & 4000 & 1000 & 0.5961 & 0.6701 & 0.7402 & 0.8067 & \textbf{4165} \\
6 & bge-base   & 0.35/0.65 & 4000 & 1000 & \textbf{0.6078} & 0.6826 & 0.7490 & 0.8238 & 4092 \\
7 & bge-base   & 0.45/0.55 & 4000 & 1000 & 0.6069 & 0.6804 & 0.7475 & 0.8171 & 4092 \\
8 & bge-base   & 0.30/0.70 & 4000 & 1500 & 0.6067 & 0.6826 & \textbf{0.7493} & \textbf{0.8284} & 4119 \\
\bottomrule
\end{tabular}
\end{table}

The results reveal a consistent trade-off between top-rank precision and recall depth. The candidate pool size, governed by the embedder and by $k_1$/$k_2$, controls how many relevant documents are recovered (\textbf{rel\_ret}) but not where they land: the wider pools retrieve more relevant documents and achieve higher NDCG@100 and reciprocal rank, yet do not improve MAP. The additional candidates surface relevant documents deep in the ranking while diluting the top ranks with false positives that the judge may incorrectly promote, because relevant documents are sparse relative to the thousands of candidates scored, MAP is dominated by precision in the top ranks.

The CE/judge blend weight reveals a dissociation between MAP and reciprocal rank. Across Runs~6-8, increasing the judge weight from $0.45/0.55$ to $0.30/0.70$ monotonically raises RR ($0.8171 \to 0.8238 \to 0.8284$), confirming that the judge excels at surfacing the single most relevant document per query. MAP, however, peaks at Run~6 ($0.6078$) and falls slightly at Run~8 ($0.6067$), indicating that beyond a $0.35/0.65$ blend the judge introduces noise in the ordering of lower-ranked candidates. The blend weight acts as a first-rank precision lever: more judge weight improves RR at a small MAP cost, and the optimal operating point depends on which metric the evaluation prioritises. For MAP-focused evaluation, Run~6 is the optimal configuration; for RR-focused evaluation, Run~8 is preferable. Overall, the spread across all configurations remains small relative to the effect of judge selection in Table~\ref{tab:judge_comparison}: the Stage~1 embedder and pool size primarily govern recall depth, while the blend weight finetunes the balance between MAP and RR within already-retrieved candidates.

\subsection{Post-Processing Grid Search} \label{ssec:post_processing_grid_search}
Having fixed the retrieval and reranking components based on Subsection~\ref{ssec:piepline_ablation}, we tune the Stage~3 scoring parameters in two passes, using the three most performant judges from Subsection~\ref{ssec:judge_comparation} (\texttt{Judge\_Qwen7B\_generic}, \texttt{Judge\_G4\_31B\_generic}, and \texttt{Judge\_G4\_31B\_typed}), cached for efficiency.

A first grid sweeps three blend values $w_{\text{CE}}/w_{\text{J}} \in \{0.30/0.70,\,0.35/0.65,\,0.40/0.60\}$, four penalty thresholds $\tau \in \{0.25, 0.30, 0.35, 0.40\}$, five penalty factors $\lambda \in \{0.70, 0.75, 0.80, 0.85, 1.00\}$, and five symmetric judge-ensemble configurations ($300$ combinations). Sweeping all 20 $\tau$-$\lambda$ combinations at fixed blend and ensemble produces identical MAP, NDCG, and RR, indicating that the judge's probability distribution is sufficiently polarized that no practical threshold separates borderline positives from negatives; we therefore fix $\tau=0.30$ and $\lambda=0.85$ and drop these dimensions.

\begin{table}[h!]
\centering
\caption{Post-processing grid search results ($\tau=0.30$, $\lambda=0.85$).
G4 $w$ is the weight of each Gemma judge (G4\_31B\_general and G4\_31B\_typed, equal).
Best values per metric are \textbf{bolded}.}
\label{tab:grid_search}
\setlength{\tabcolsep}{4pt}
\begin{tabular}{llccccc}
\toprule
$w_{\text{CE}}$/$w_{\text{J}}$ & \textbf{Qwen} $w$ & \textbf{G4} $w$ & \textbf{MAP} & \textbf{NDCG@10} & \textbf{NDCG@100} & \textbf{recip\_rank (RR)} \\
\midrule
0.30/0.70 & 1.00 & 0.00 & 0.6049 & 0.6812 & 0.7477 & 0.8263 \\
0.35/0.65 & 1.00 & 0.00 & 0.6078 & 0.6826 & 0.7490 & 0.8238 \\
0.40/0.60 & 1.00 & 0.00 & 0.6072 & 0.6819 & 0.7485 & 0.8200 \\
\midrule
0.30/0.70 & 0.60 & 0.20 & \textbf{0.6343} & \textbf{0.7049} & \textbf{0.7680} & \textbf{0.8267} \\
0.35/0.65 & 0.60 & 0.20 & 0.6327 & 0.7011 & 0.7661 & 0.8225 \\
0.40/0.60 & 0.60 & 0.20 & 0.6315 & 0.7004 & 0.7654 & 0.8217 \\
\midrule
0.30/0.70 & 0.50 & 0.25 & 0.6329 & 0.7020 & 0.7668 & 0.8216 \\
0.35/0.65 & 0.50 & 0.25 & 0.6335 & 0.7019 & 0.7672 & 0.8220 \\
0.40/0.60 & 0.50 & 0.25 & 0.6304 & 0.6999 & 0.7646 & 0.8194 \\
\midrule
0.30/0.70 & 0.40 & 0.30 & 0.6339 & 0.7016 & 0.7674 & 0.8223 \\
0.35/0.65 & 0.40 & 0.30 & 0.6305 & 0.6994 & 0.7650 & 0.8208 \\
0.40/0.60 & 0.40 & 0.30 & 0.6298 & 0.6885 & 0.7641 & 0.8192 \\
\midrule
0.30/0.70 & 0.33 & 0.33 & 0.6308 & 0.6988 & 0.7651 & 0.8192 \\
0.35/0.65 & 0.33 & 0.33 & 0.6296 & 0.6986 & 0.7643 & 0.8191 \\
0.40/0.60 & 0.33 & 0.33 & 0.6285 & 0.6966 & 0.7631 & 0.8166 \\
\bottomrule
\end{tabular}
\end{table}

Table~\ref{tab:grid_search} reports the blend $\times$ symmetric-ensemble results. Adding Gemma judges at equal weights beats the Qwen-only baseline, confirming that the Qwen and Gemma judges are decorrelated enough to provide complementary ranking signals. The best symmetric configuration is Qwen~$0.60$ with each Gemma judge at $0.20$ and blend $0.30/0.70$, reaching MAP $0.6343$; reducing the Qwen weight further slightly degrades the metrics, so Qwen remains the primary signal and the Gemma judges provide complementary corrections rather than an equal vote. The $0.30/0.70$ blend becomes optimal with the ensemble (versus $0.35/0.65$ for the Qwen-only configuration), which we attribute to the ensemble producing a better-calibrated, more reliable signal that tolerates a higher judge weight without the noise penalty seen with a single model.

A second, finer search relaxes the equal-Gemma constraint, jointly sweeping the Qwen weight $w_{\text{Qwen}}$ from $0.50$ to $0.95$ (step $0.05$) against an asymmetric split of the remaining budget between \texttt{Judge\_G4\_31B\_generic} and \texttt{Judge\_G4\_31B\_typed} ($w_{\text{G4generic}}$ from $0$ to $b$ in steps of $0.05$, with $w_{\text{G4typed}} = b - w_{\text{G4generic}}$), crossed with three blends $w_{\text{CE}}/w_{\text{J}} \in \{0.25/0.75,\,0.30/0.70,\,0.35/0.65\}$, for $195$ submissions. Table~\ref{tab:expanded_grid_search} reports the configurations at or above the symmetric baseline. Our submitted configuration (Qwen~$0.60$, Gemma-generic~$0.30$, Gemma-typed~$0.10$, blend~$0.30/0.70$) reaches MAP $0.6347$, but the top entries are effectively tied ($0.6343$-$0.6347$): the asymmetric tuning confirms a plateau rather than yielding a further gain. The $0.30/0.70$ blend is the most stable across all triples, the \texttt{G4generic}/\texttt{G4typed} split has minimal impact on MAP (above-baseline configurations span G4generic weights $0.10$-$0.30$, so the two Gemma judges are largely interchangeable), and \texttt{rel\_ret} is flat at $4092$, confirming that ensemble weight tuning affects only precision ordering and not recall depth. The ensemble weight space is therefore exhausted under the fixed Stage~1 and Stage~2 components.

\begin{table}[h!]
\centering
\caption{%
  Expanded asymmetric ensemble grid search results ($\tau=0.30$, $\lambda=0.85$, $w_{\text{CE}}$/$w_{\text{J}}$ = $0.30/0.70$, $w_{\text{Qwen}}$ = $0.60$). Only configurations with MAP~$> 0.6343$ (previous best) are shown.
  Best values per metric are \textbf{bolded}.%
}
\label{tab:expanded_grid_search}
\setlength{\tabcolsep}{4pt}
\begin{tabular}{lcccccccc}
\toprule
$w_{\text{G4general}}$ &
$w_{\text{G4typed}}$ &
\textbf{MAP} &
\textbf{NDCG@5} &
\textbf{NDCG@10} &
\textbf{NDCG@100} &
\textbf{recip\_rank (RR)} &
\textbf{P@10} &
\textbf{rel\_ret} \\
\midrule
 0.30 & 0.10 & \textbf{0.6347} & \textbf{0.6924} & 0.7050 & \textbf{0.7684} & \textbf{0.8273} & 0.3124 & 4092 \\
 0.10 & 0.30 & \textbf{0.6347} & 0.6916 & 0.7048 & 0.7682 & 0.8268 & 0.3126 & 4092 \\
 0.25 & 0.15 & 0.6345 & 0.6915 & \textbf{0.7051} & 0.7682 & \textbf{0.8273} & \textbf{0.3127} & 4092 \\
 0.15 & 0.25 & 0.6344 & 0.6918 & 0.7044 & 0.7681 & 0.8268 & 0.3122 & 4092 \\
\midrule
 0.20 & 0.20 & 0.6343 & 0.6916 & 0.7049 & 0.7680 & 0.8267 & 0.3127 & 4092 \\
\bottomrule
\end{tabular}
\end{table}

\subsection{Discussions}\label{ssec:discussion}
A single theme runs through our pipeline ablation: despite the theoretical advantages of larger, more expensive components, lighter configurations consistently outperformed their heavier counterparts on the official evaluation. For context against prior editions of the task, the best official MAP reported in the JOKER~2024 Task~1 overview was $0.12$~\cite{ermakova2024overview}, while the strongest 2025 English run reached $0.3501$, obtained by a Qwen-based filter-explainer with dense retrieval that outperformed the next-best team by roughly a factor of two~\cite{ermakova2025overview,vachharajani2025pjmathematician}. IROH's $0.6347$ thus represents a substantial advance over the best previously reported systems on JOKER Task~1 English, and extends the LLM-filtering line of work that produced the 2025 state of the art. We note that the corpus has been expanded across editions, so this comparison is indicative of progress on the task rather than a strictly controlled benchmark. The smaller \texttt{bge-base-en-v1.5} embedder (768-dim) achieved a higher MAP than \texttt{bge-en-icl} (4096-dim, 7B parameters) when paired with the full judge pipeline, and the best Qwen2.5-7B judge, trained on generic rationales, outperformed every Gemma-4-31B configuration, although on the typed rationales Gemma-4-31B was the stronger of the two. We hypothesize that this is partly a calibration effect: lighter models produce smoother score distributions that blend more effectively with the cross-encoder signal, while larger models tend toward more binary outputs that are harder to control in a multi-signal fusion setting. This suggests that for humor retrieval, where the distinction between relevant and non-relevant texts is subtle and domain-specific, finetuning quality and training data composition matter more than raw model capacity.

A parallel finding concerns the two rationale strategies. The generic pipeline, which uses a single lightweight placeholder as query context, consistently outperforms the typed pipeline, which uses a richer structured prompt enumerating seven humor mechanism types. We hypothesize that forcing the model to classify each text into a specific humor category introduces variance in rationale quality, since Gemma~4 may not reliably distinguish between, for example, homophonic and homographic puns in short texts. The generic pipeline avoids this by asking only for a one-sentence explanation of why the text is or is not funny, producing more consistent supervision signal across training examples.

A further structural finding is that the recall bottleneck lies in Stage~1, not Stage~3. The \texttt{rel\_ret} metric is flat at $4092$ across all ensemble configurations in the expanded grid search, but increases from $3564$ to $4092$ when the candidate pool is widened from $k_1 = 3000$ to $k_1 = 4000$ in the pipeline ablation. This confirms that the judge ensemble can only rerank what Stage~1 retrieves, and that future gains in recall depth require improvements to the first-stage retrieval rather than further tuning of the ensemble weights.

A final ablation isolates the rationales' contribution to retrieval. Running the pipeline with the judge disabled (Stages~1-2 only) on the official test queries, we compared a plain BM25 index against one expanded with the document rationales while holding the cross-encoder fixed. The effect was negligible: rationale-expanded indexing reached a MAP of $0.2876$ against $0.2868$ for plain BM25, with NDCG@10 ($0.3485$ vs.\ $0.3491$) and the number of relevant documents retrieved ($3913$ vs.\ $3918$) essentially unchanged, and the sign of the difference flipping across metrics. The rationales thus contribute little at the lexical-retrieval level; within our pipeline, they enter primarily as the supervision signal for the distilled judge rather than as an indexing aid. This is consistent with our finding that the judge, not first-stage retrieval, is the dominant top-rank signal, and it is a further instance of the local-validation signal, where the same comparison showed a much larger gap, overstating an effect that does not survive on the official benchmark.

\subsection{Limitations}
Our pipeline is resource and time-intensive, which constrained the breadth of our experimental evaluation. The full pipeline, including rationale generation, cross-encoder finetuning, and judge finetuning, requires multiple days of compute on multiple GPUs, and CodaBench submission limits further constrained the hyperparameter search, requiring us to rely on OFAT sweeps rather than principled grid search for several components.

The generated rationales, while useful as a training signal, have a ceiling in quality. For straightforward wordplay, such as the "fortitude" Tom Swifty example, the model produces accurate and specific rationales. But for humor that relies on cultural context, implicit shared knowledge, or multi-layered irony, the rationale generation tends to produce generic explanations that do not capture the actual mechanism of the joke, limiting the quality of the supervision signal for these cases.

A further limitation is the binary YES/NO framing of the judge. Humor relevance is arguably a spectrum: a text can be partially relevant, tangentially funny, or topically related but not humorous in the expected way. The binary verdict forces a hard decision that may be inappropriate for borderline cases, and a soft relevance score trained on graded judgments could produce better-calibrated judge outputs.

Finally, our pipeline is English-only. The rationale generation prompts, the humor mechanism taxonomy, and all finetuned components are trained exclusively on English data, and it is unclear how well the approach would transfer to the other languages in the JOKER corpus without retraining.

\section{Conclusions} \label{sec:conclusions}
We presented IROH, a three-stage retrieval system for humor-aware information retrieval that combines hybrid sparse-dense retrieval, cross-encoder reranking, and a weighted ensemble of LoRA-adapted LLM judges, achieving a best MAP of $0.6347$ and \textbf{ranking first on} the official JOKER~2026 Task~1 English leaderboard.

Our ablation study yields three findings of broader relevance: the rationale-distilled judge is the primary driver of ranking quality, while rationale-expanded first-stage indexing has a negligible effect on retrieval (a MAP difference of $0.0008$ on the official test); structured hard negatives degrade generalisation in nearly all configurations despite inflating local validation scores, an effect we attribute to distribution shift between the local split and the broader retrieval task; and within our pipeline the lighter, better-calibrated model matches or beats its larger counterpart, suggesting that for this task finetuning quality and data composition matter at least as much as raw model capacity for subtle, domain-specific tasks such as humor retrieval.

A parallel finding on rationale strategy shows that the simpler generic pipeline outperforms the richer typed variant, which we attribute to variance in rationale quality introduced by forcing the model to classify humor into specific mechanism types. An extended ensemble grid search over $195$ configurations confirmed that the weight space is effectively exhausted under the fixed Stage~1 and Stage~2 components, with \texttt{rel\_ret} flat across all ensemble runs, indicating that further gains in recall require improvements to first-stage retrieval rather than reranking precision.

For future work, we plan to explore multilingual humor retrieval as a general solution across all JOKER Task~1 languages. On the retrieval side, the rel\_ret analysis suggests that improving first-stage recall, through better query expansion or domain-adapted dense encoders, would yield larger gains than any further ensemble tuning. On the model side, the calibration hypothesis identified in our discussion suggests that better rationale quality, particularly for humor relying on cultural context or implicit shared knowledge, would benefit the judge's decision process more than increasing model capacity. Finally, replacing the binary YES/NO judge framing with graded relevance training remains an open and promising direction for producing better-calibrated judge outputs.

\begin{acknowledgments}
The research presented in this paper was supported in part by (1) The Academy of Romanian Scientists, through the funding of the project ``NetGuardAI: Intelligent system for harmful content detection and immunization on social networks'' (AOȘR-TEAMS-IV); and
(2) the National University of Science and Technology, POLITEHNICA Bucharest, through the PubArt program.
\end{acknowledgments}

\section*{Declaration on Generative AI}
  During the preparation of this work, the authors used Claude Opus 4.6 for rephrasing in order to improve clarity and style. After using this tool, the authors reviewed and edited the content as needed and take full responsibility for the publication’s content.
  

\bibliography{joker}

@article{suls1972two,
  title={A two-stage model for the appreciation of jokes and cartoons: An information-processing analysis},
  author={Suls, Jerry M},
  journal={The psychology of humor: Theoretical perspectives and empirical issues},
  volume={1},
  pages={81--100},
  year={1972}
}

@inproceedings{weller2019humor,
    title = "Humor Detection: A Transformer Gets the Last Laugh",
    author = "Weller, Orion  and
      Seppi, Kevin",
    editor = "Inui, Kentaro  and
      Jiang, Jing  and
      Ng, Vincent  and
      Wan, Xiaojun",
    booktitle = "Proceedings of the 2019 Conference on Empirical Methods in Natural Language Processing and the 9th International Joint Conference on Natural Language Processing (EMNLP-IJCNLP)",
    month = nov,
    year = "2019",
    address = "Hong Kong, China",
    publisher = "Association for Computational Linguistics",
    url = "https://aclanthology.org/D19-1372/",
    doi = "10.18653/v1/D19-1372",
    pages = "3621--3625"
}

@inproceedings{xie2021uncertainty,
  title = "Uncertainty and Surprisal Jointly Deliver the Punchline: Exploiting Incongruity-Based Features for Humor Recognition",
    author = "Xie, Yubo  and
      Li, Junze  and
      Pu, Pearl",
    editor = "Zong, Chengqing  and
      Xia, Fei  and
      Li, Wenjie  and
      Navigli, Roberto",
    booktitle = "Proceedings of the 59th Annual Meeting of the Association for Computational Linguistics and the 11th International Joint Conference on Natural Language Processing (Volume 2: Short Papers)",
    month = aug,
    year = "2021",
    address = "Online",
    publisher = "Association for Computational Linguistics",
    url = "https://aclanthology.org/2021.acl-short.6/",
    doi = "10.18653/v1/2021.acl-short.6",
    pages = "33--39"
}

@inproceedings{ermakova2024overview,
  author       = {Liana Ermakova and
                  Anne{-}Gwenn Bosser and
                  Tristan Miller and
                  Adam Jatowt},
  editor       = {Guglielmo Faggioli and
                  Nicola Ferro and
                  Petra Galusc{\'{a}}kov{\'{a}} and
                  Alba Garc{\'{\i}}a Seco de Herrera},
  title        = {Overview of the {CLEF} 2024 {JOKER} Task 1: Humour-aware Information
                  Retrieval},
  booktitle    = {Working Notes of the Conference and Labs of the Evaluation Forum {(CLEF}
                  2024), Grenoble, France, 9-12 September, 2024},
  series       = {{CEUR} Workshop Proceedings},
  pages        = {1775--1785},
  publisher    = {CEUR-WS.org},
  year         = {2024},
  url          = {https://ceur-ws.org/Vol-3740/paper-165.pdf},
  bibsource    = {dblp computer science bibliography, https://dblp.org}
}

@inproceedings{schuurman2024university,
  author       = {Emma Schuurman and
                  Mick Cazemier and
                  Luc Buijs and
                  Jaap Kamps},
  editor       = {Guglielmo Faggioli and
                  Nicola Ferro and
                  Petra Galusc{\'{a}}kov{\'{a}} and
                  Alba Garc{\'{\i}}a Seco de Herrera},
  title        = {University of Amsterdam at the {CLEF} 2024 Joker Track},
  booktitle    = {Working Notes of the Conference and Labs of the Evaluation Forum {(CLEF}
                  2024), Grenoble, France, 9-12 September, 2024},
  series       = {{CEUR} Workshop Proceedings},
  pages        = {1909--1922},
  publisher    = {CEUR-WS.org},
  year         = {2024},
  url          = {https://ceur-ws.org/Vol-3740/paper-181.pdf},
  bibsource    = {dblp computer science bibliography, https://dblp.org}
}

@inproceedings{vachharajani2025pjmathematician,
  author       = {Poojan Vachharajani},
  editor       = {Guglielmo Faggioli and
                  Nicola Ferro and
                  Paolo Rosso and
                  Damiano Spina},
  title        = {pjmathematician at the {CLEF} 2025 {JOKER} Lab Tasks 1, 2 {\&}
                  3: {A} Unified Approach to Humour Retrieval and Translation using
                  the Qwen {LLM} Family},
  booktitle    = {Working Notes of the Conference and Labs of the Evaluation Forum,
                  {CLEF} 2025, Madrid, Spain, 9-12 September 2025},
  series       = {{CEUR} Workshop Proceedings},
  pages        = {2889--2897},
  publisher    = {CEUR-WS.org},
  year         = {2025},
  url          = {https://ceur-ws.org/Vol-4038/paper\_230.pdf},
  bibsource    = {dblp computer science bibliography, https://dblp.org}
}

@inproceedings{zhuang2023rankt5,
  author       = {Honglei Zhuang and
                  Zhen Qin and
                  Rolf Jagerman and
                  Kai Hui and
                  Ji Ma and
                  Jing Lu and
                  Jianmo Ni and
                  Xuanhui Wang and
                  Michael Bendersky},
  editor       = {Hsin{-}Hsi Chen and
                  Wei{-}Jou (Edward) Duh and
                  Hen{-}Hsen Huang and
                  Makoto P. Kato and
                  Josiane Mothe and
                  Barbara Poblete},
  title        = {RankT5: Fine-Tuning {T5} for Text Ranking with Ranking Losses},
  booktitle    = {Proceedings of the 46th International {ACM} {SIGIR} Conference on
                  Research and Development in Information Retrieval, {SIGIR} 2023, Taipei,
                  Taiwan, July 23-27, 2023},
  pages        = {2308--2313},
  publisher    = {{ACM}},
  year         = {2023},
  url          = {https://doi.org/10.1145/3539618.3592047},
  doi          = {10.1145/3539618.3592047},
  bibsource    = {dblp computer science bibliography, https://dblp.org}
}

@inproceedings{hsieh2023distilling,
    title = "Distilling Step-by-Step! Outperforming Larger Language Models with Less Training Data and Smaller Model Sizes",
    author = "Hsieh, Cheng-Yu  and
      Li, Chun-Liang  and
      Yeh, Chih-kuan  and
      Nakhost, Hootan  and
      Fujii, Yasuhisa  and
      Ratner, Alex  and
      Krishna, Ranjay  and
      Lee, Chen-Yu  and
      Pfister, Tomas",
    editor = "Rogers, Anna  and
      Boyd-Graber, Jordan  and
      Okazaki, Naoaki",
    booktitle = "Findings of the Association for Computational Linguistics: ACL 2023",
    month = jul,
    year = "2023",
    address = "Toronto, Canada",
    publisher = "Association for Computational Linguistics",
    url = "https://aclanthology.org/2023.findings-acl.507/",
    doi = "10.18653/v1/2023.findings-acl.507",
    pages = "8003--8017"
}

@inproceedings{tian2025beyond,
   author       = {Yijun Tian and
                  Yikun Han and
                  Xiusi Chen and
                  Wei Wang and
                  Nitesh V. Chawla},
  editor       = {Wolfgang Nejdl and
                  S{\"{o}}ren Auer and
                  Meeyoung Cha and
                  Marie{-}Francine Moens and
                  Marc Najork},
  title        = {Beyond Answers: Transferring Reasoning Capabilities to Smaller LLMs
                  Using Multi-Teacher Knowledge Distillation},
  booktitle    = {Proceedings of the Eighteenth {ACM} International Conference on Web
                  Search and Data Mining, {WSDM} 2025, Hannover, Germany, March 10-14,
                  2025},
  pages        = {251--260},
  publisher    = {{ACM}},
  year         = {2025},
  url          = {https://doi.org/10.1145/3701551.3703577},
  doi          = {10.1145/3701551.3703577},
  bibsource    = {dblp computer science bibliography, https://dblp.org}
}

@inproceedings{li2025making,
  author       = {Chaofan Li and
                  Minghao Qin and
                  Shitao Xiao and
                  Jianlyu Chen and
                  Kun Luo and
                  Defu Lian and
                  Yingxia Shao and
                  Zheng Liu},
  title        = {Making Text Embedders Few-Shot Learners},
  booktitle    = {The Thirteenth International Conference on Learning Representations,
                  {ICLR} 2025, Singapore, April 24-28, 2025},
  publisher    = {OpenReview.net},
  year         = {2025},
  url          = {https://openreview.net/forum?id=wfLuiDjQ0u},
  bibsource    = {dblp computer science bibliography, https://dblp.org}
}

@inproceedings{cormack2009reciprocal,
  author       = {Gordon V. Cormack and
                  Charles L. A. Clarke and
                  Stefan B{\"{u}}ttcher},
  editor       = {James Allan and
                  Javed A. Aslam and
                  Mark Sanderson and
                  ChengXiang Zhai and
                  Justin Zobel},
  title        = {Reciprocal rank fusion outperforms condorcet and individual rank learning
                  methods},
  booktitle    = {Proceedings of the 32nd Annual International {ACM} {SIGIR} Conference
                  on Research and Development in Information Retrieval, {SIGIR} 2009,
                  Boston, MA, USA, July 19-23, 2009},
  pages        = {758--759},
  publisher    = {{ACM}},
  year         = {2009},
  url          = {https://doi.org/10.1145/1571941.1572114},
  doi          = {10.1145/1571941.1572114},
  bibsource    = {dblp computer science bibliography, https://dblp.org}
}

@inproceedings{romanowski2025punchlines,
title = "From Punchlines to Predictions: A Metric to Assess {LLM} Performance in Identifying Humor in Stand-Up Comedy",
    author = "Romanowski, Adrianna  and
      Valois, Pedro H. V.  and
      Fukui, Kazuhiro",
    editor = "Kuribayashi, Tatsuki  and
      Rambelli, Giulia  and
      Takmaz, Ece  and
      Wicke, Philipp  and
      Li, Jixing  and
      Oh, Byung-Doh",
    booktitle = "Proceedings of the Workshop on Cognitive Modeling and Computational Linguistics",
    month = may,
    year = "2025",
    address = "Albuquerque, New Mexico, USA",
    publisher = "Association for Computational Linguistics",
    url = "https://aclanthology.org/2025.cmcl-1.6/",
    doi = "10.18653/v1/2025.cmcl-1.6",
    pages = "36--46",
    ISBN = "979-8-89176-227-5"
}

@inproceedings{ermakova2025overview,
  author       = {Liana Ermakova and
                  Ricardo Campos and
                  Anne{-}Gwenn Bosser and
                  Tristan Miller},
  editor       = {Jorge Carrillo{-}de{-}Albornoz and
                  Alba Garc{\'{\i}}a Seco de Herrera and
                  Julio Gonzalo and
                  Laura Plaza and
                  Josiane Mothe and
                  Florina Piroi and
                  Paolo Rosso and
                  Damiano Spina and
                  Guglielmo Faggioli and
                  Nicola Ferro},
  title        = {Overview of the {CLEF} 2025 {JOKER} Lab: Humour in Machine},
  booktitle    = {Experimental {IR} Meets Multilinguality, Multimodality, and Interaction
                  - 16th International Conference of the {CLEF} Association, {CLEF}
                  2025, Madrid, Spain, September 9-12, 2025, Proceedings},
  series       = {Lecture Notes in Computer Science},
  pages        = {315--337},
  publisher    = {Springer},
  year         = {2025},
  url          = {https://doi.org/10.1007/978-3-032-04354-2\_18},
  doi          = {10.1007/978-3-032-04354-2\_18},
  bibsource    = {dblp computer science bibliography, https://dblp.org}
}

@inproceedings{reimers2019sentence,
  author       = {Nils Reimers and
                  Iryna Gurevych},
  editor       = {Kentaro Inui and
                  Jing Jiang and
                  Vincent Ng and
                  Xiaojun Wan},
  title        = {Sentence-BERT: Sentence Embeddings using Siamese BERT-Networks},
  booktitle    = {Proceedings of the 2019 Conference on Empirical Methods in Natural
                  Language Processing and the 9th International Joint Conference on
                  Natural Language Processing, {EMNLP-IJCNLP} 2019, Hong Kong, China,
                  November 3-7, 2019},
  pages        = {3980--3990},
  publisher    = {Association for Computational Linguistics},
  year         = {2019},
  url          = {https://doi.org/10.18653/v1/D19-1410},
  doi          = {10.18653/V1/D19-1410},
  bibsource    = {dblp computer science bibliography, https://dblp.org}
}

@inproceedings{zhang2024mgte,
    title = "{mGTE}: Generalized Long-Context Text Representation and Reranking Models for Multilingual Text Retrieval",
    author = "Zhang, Xin  and
      Zhang, Yanzhao  and
      Long, Dingkun  and
      Xie, Wen  and
      Dai, Ziqi  and
      Tang, Jialong  and
      Lin, Huan  and
      Yang, Baosong  and
      Xie, Pengjun  and
      Huang, Fei  and
      Zhang, Meishan  and
      Li, Wenjie  and
      Zhang, Min",
    editor = "Dernoncourt, Franck  and
      Preo{\c{t}}iuc-Pietro, Daniel  and
      Shimorina, Anastasia",
    booktitle = "Proceedings of the 2024 Conference on Empirical Methods in Natural Language Processing: Industry Track",
    month = nov,
    year = "2024",
    address = "Miami, Florida, US",
    publisher = "Association for Computational Linguistics",
    url = "https://aclanthology.org/2024.emnlp-industry.103/",
    doi = "10.18653/v1/2024.emnlp-industry.103",
    pages = "1393--1412"
}

@article{sinha2025don,
  author       = {Aarush Sinha},
  title        = {Don't Retrieve, Generate: Prompting LLMs for Synthetic Training
                  Data in Dense Retrieval},
  journal      = {CoRR},
  volume       = {abs/2504.21015},
  year         = {2025},
  url          = {https://doi.org/10.48550/arXiv.2504.21015},
  doi          = {10.48550/ARXIV.2504.21015},
  eprinttype   = {arXiv},
  eprint       = {2504.21015},
  bibsource    = {dblp computer science bibliography, https://dblp.org}
}

@inproceedings{meghwani2025hard,
title = "Hard Negative Mining for Domain-Specific Retrieval in Enterprise Systems",
    author = "Meghwani, Hansa  and
      Agarwal, Amit  and
      Pattnayak, Priyaranjan  and
      Patel, Hitesh Laxmichand  and
      Panda, Srikant",
    editor = "Rehm, Georg  and
      Li, Yunyao",
    booktitle = "Proceedings of the 63rd Annual Meeting of the Association for Computational Linguistics (Volume 6: Industry Track)",
    month = jul,
    year = "2025",
    address = "Vienna, Austria",
    publisher = "Association for Computational Linguistics",
    url = "https://aclanthology.org/2025.acl-industry.72/",
    doi = "10.18653/v1/2025.acl-industry.72",
    pages = "1013--1026",
    ISBN = "979-8-89176-288-6"
}

@inproceedings{balog2025rankers,
  author       = {Krisztian Balog and
                  Don Metzler and
                  Zhen Qin},
  editor       = {Nicola Ferro and
                  Maria Maistro and
                  Gabriella Pasi and
                  Omar Alonso and
                  Andrew Trotman and
                  Suzan Verberne},
  title        = {Rankers, Judges, and Assistants: Towards Understanding the Interplay
                  of LLMs in Information Retrieval Evaluation},
  booktitle    = {Proceedings of the 48th International {ACM} {SIGIR} Conference on
                  Research and Development in Information Retrieval, {SIGIR} 2025, Padua,
                  Italy, July 13-18, 2025},
  pages        = {3865--3875},
  publisher    = {{ACM}},
  year         = {2025},
  url          = {https://doi.org/10.1145/3726302.3730348},
  doi          = {10.1145/3726302.3730348},
  bibsource    = {dblp computer science bibliography, https://dblp.org}
}

@inproceedings{miller2017semeval,
    title = "{S}em{E}val-2017 Task 7: Detection and Interpretation of {E}nglish Puns",
    author = "Miller, Tristan  and
      Hempelmann, Christian  and
      Gurevych, Iryna",
    editor = "Bethard, Steven  and
      Carpuat, Marine  and
      Apidianaki, Marianna  and
      Mohammad, Saif M.  and
      Cer, Daniel  and
      Jurgens, David",
    booktitle = "Proceedings of the 11th International Workshop on Semantic Evaluation ({S}em{E}val-2017)",
    month = aug,
    year = "2017",
    address = "Vancouver, Canada",
    publisher = "Association for Computational Linguistics",
    url = "https://aclanthology.org/S17-2005/",
    doi = "10.18653/v1/S17-2005",
    pages = "58--68"
}

@inproceedings{kiddon2011s,
  author       = {Chlo{\'{e}} Kiddon and
                  Yuriy Brun},
  title        = {That's What She Said: Double Entendre Identification},
  booktitle    = {The 49th Annual Meeting of the Association for Computational Linguistics:
                  Human Language Technologies, Proceedings of the Conference, 19-24
                  June, 2011, Portland, Oregon, {USA} - Short Papers},
  pages        = {89--94},
  publisher    = {The Association for Computer Linguistics},
  year         = {2011},
  url          = {https://aclanthology.org/P11-2016/},
  bibsource    = {dblp computer science bibliography, https://dblp.org}
}

@inproceedings{lessard1992computational,
  title={Computational model ling of linguistic humour: Tom swifty},
  author={Lessard, G and Levison, M},
  booktitle={Paper Delivered at the ALLC/ACH Joint Annual Conference. Christ Church, Oxford},
  year={1992}
}

@article{hui2024qwen2,
  title={Qwen2.5 Technical Report},
  author={Qwen An Yang and Baosong Yang and Beichen Zhang and Binyuan Hui and Bo Zheng and Bowen Yu and Chengyuan Li and Dayiheng Liu and Fei Huang and Guanting Dong and Haoran Wei and Huan Lin and Jian Yang and Jianhong Tu and Jianwei Zhang and Jianxin Yang and Jiaxin Yang and Jingren Zhou and Junyang Lin and Kai Dang and Keming Lu and Keqin Bao and Kexin Yang and Le Yu and Mei Li and Mingfeng Xue and Pei Zhang and Qin Zhu and Rui Men and Runji Lin and Tianhao Li and Tingyu Xia and Xingzhang Ren and Xuancheng Ren and Yang Fan and Yang Su and Yi-Chao Zhang and Yunyang Wan and Yuqi Liu and Zeyu Cui and Zhenru Zhang and Zihan Qiu and Shanghaoran Quan and Zekun Wang},
  journal={ArXiv},
  year={2024},
  volume={abs/2412.15115},
  url={https://api.semanticscholar.org/CorpusID:274859421},
  doi={https://doi.org/10.48550/arXiv.2412.15115}
}

@inproceedings{chen2024bge,
    title = "{M}3-Embedding: Multi-Linguality, Multi-Functionality, Multi-Granularity Text Embeddings Through Self-Knowledge Distillation",
    author = "Chen, Jianlyu  and
      Xiao, Shitao  and
      Zhang, Peitian  and
      Luo, Kun  and
      Lian, Defu  and
      Liu, Zheng",
    editor = "Ku, Lun-Wei  and
      Martins, Andre  and
      Srikumar, Vivek",
    booktitle = "Findings of the Association for Computational Linguistics: ACL 2024",
    month = aug,
    year = "2024",
    address = "Bangkok, Thailand",
    publisher = "Association for Computational Linguistics",
    url = "https://aclanthology.org/2024.findings-acl.137/",
    doi = "10.18653/v1/2024.findings-acl.137",
    pages = "2318--2335"
}

@inproceedings{xiao2024cpack,
  author       = {Shitao Xiao and
                  Zheng Liu and
                  Peitian Zhang and
                  Niklas Muennighoff and
                  Defu Lian and
                  Jian{-}Yun Nie},
  editor       = {Grace Hui Yang and
                  Hongning Wang and
                  Sam Han and
                  Claudia Hauff and
                  Guido Zuccon and
                  Yi Zhang},
  title        = {C-Pack: Packed Resources For General Chinese Embeddings},
  booktitle    = {Proceedings of the 47th International {ACM} {SIGIR} Conference on
                  Research and Development in Information Retrieval, {SIGIR} 2024, Washington
                  DC, USA, July 14-18, 2024},
  pages        = {641--649},
  publisher    = {{ACM}},
  year         = {2024},
  url          = {https://doi.org/10.1145/3626772.3657878},
  doi          = {10.1145/3626772.3657878},
  bibsource    = {dblp computer science bibliography, https://dblp.org}
}

@article{nogueira2019passage,
  author       = {Rodrigo Nogueira and
                  Kyunghyun Cho},
  title        = {Passage Re-ranking with {BERT}},
  journal      = {CoRR},
  volume       = {abs/1901.04085},
  year         = {2019},
  url          = {http://arxiv.org/abs/1901.04085},
  eprinttype   = {arXiv},
  eprint       = {1901.04085},
  bibsource    = {dblp computer science bibliography, https://dblp.org}
}

@inproceedings{xiong2020approximate,
  author       = {Lee Xiong and
                  Chenyan Xiong and
                  Ye Li and
                  Kwok{-}Fung Tang and
                  Jialin Liu and
                  Paul N. Bennett and
                  Junaid Ahmed and
                  Arnold Overwijk},
  title        = {Approximate Nearest Neighbor Negative Contrastive Learning for Dense
                  Text Retrieval},
  booktitle    = {9th International Conference on Learning Representations, {ICLR} 2021,
                  Virtual Event, Austria, May 3-7, 2021},
  publisher    = {OpenReview.net},
  year         = {2021},
  url          = {https://openreview.net/forum?id=zeFrfgyZln},
  bibsource    = {dblp computer science bibliography, https://dblp.org}
}

@article{li2024syneg,
  author       = {Xiaopeng Li and
                  Xiangyang Li and
                  Hao Zhang and
                  Zhaocheng Du and
                  Pengyue Jia and
                  Yichao Wang and
                  Xiangyu Zhao and
                  Huifeng Guo and
                  Ruiming Tang},
  title        = {SyNeg: LLM-Driven Synthetic Hard-Negatives for Dense Retrieval},
  journal      = {CoRR},
  volume       = {abs/2412.17250},
  year         = {2024},
  url          = {https://doi.org/10.48550/arXiv.2412.17250},
  doi          = {10.48550/ARXIV.2412.17250},
  eprinttype   = {arXiv},
  eprint       = {2412.17250},
  bibsource    = {dblp computer science bibliography, https://dblp.org}
}

@inproceedings{rahmani2024llmjudge,
  author       = {Hossein A. Rahmani and
                  Emine Yilmaz and
                  Nick Craswell and
                  Bhaskar Mitra and
                  Paul Thomas and
                  Charles L. A. Clarke and
                  Mohammad Aliannejadi and
                  Clemencia Siro and
                  Guglielmo Faggioli},
  editor       = {Clemencia Siro and
                  Mohammad Aliannejadi and
                  Hossein A. Rahmani and
                  Nick Craswell and
                  Charles L. A. Clarke and
                  Guglielmo Faggioli and
                  Bhaskar Mitra and
                  Paul Thomas and
                  Emine Yilmaz},
  title        = {LLMJudge: LLMs for Relevance Judgments},
  booktitle    = {Proceedings of The First Workshop on Large Language Models for Evaluation
                  in Information Retrieval (LLM4Eval 2024) co-located with 10th International
                  Conference on Online Publishing {(SIGIR} 2024), Washington D.C., USA,
                  July 18, 2024},
  series       = {{CEUR} Workshop Proceedings},
  pages        = {1--3},
  publisher    = {CEUR-WS.org},
  year         = {2024},
  url          = {https://ceur-ws.org/Vol-3752/paper8.pdf},
  bibsource    = {dblp computer science bibliography, https://dblp.org}
}

@article{niu2024judgerank,
  author       = {Tong Niu and
                  Shafiq Joty and
                  Ye Liu and
                  Caiming Xiong and
                  Yingbo Zhou and
                  Semih Yavuz},
  title        = {JudgeRank: Leveraging Large Language Models for Reasoning-Intensive
                  Reranking},
  journal      = {CoRR},
  volume       = {abs/2411.00142},
  year         = {2024},
  url          = {https://doi.org/10.48550/arXiv.2411.00142},
  doi          = {10.48550/ARXIV.2411.00142},
  eprinttype   = {arXiv},
  eprint       = {2411.00142},
  bibsource    = {dblp computer science bibliography, https://dblp.org}
}

@inproceedings{dettmers2023qlora,
  author       = {Tim Dettmers and
                  Artidoro Pagnoni and
                  Ari Holtzman and
                  Luke Zettlemoyer},
  editor       = {Alice Oh and
                  Tristan Naumann and
                  Amir Globerson and
                  Kate Saenko and
                  Moritz Hardt and
                  Sergey Levine},
  title        = {QLoRA: Efficient Finetuning of Quantized LLMs},
  booktitle    = {Advances in Neural Information Processing Systems 36: Annual Conference
                  on Neural Information Processing Systems 2023, NeurIPS 2023, New Orleans,
                  LA, USA, December 10 - 16, 2023},
  year         = {2023},
  url          = {http://papers.nips.cc/paper\_files/paper/2023/hash/1feb87871436031bdc0f2beaa62a049b-Abstract-Conference.html},
  bibsource    = {dblp computer science bibliography, https://dblp.org}
}

@inproceedings{joker-1,
  author    = {Poojan Vachharajani and others},
  title     = {{Overview of the CLEF JOKER Task 1: Humor-Aware Information Retrieval in English and Hinglish}},
  crossref  = {clef:ceur26},
}

@inproceedings{joker-lncs,
  author    = {Liana Ermakova and others},
  title     = {Overview of the {CLEF 2026 JOKER} Track: Humor Detection, Search, and Translation},
  crossref  = {clef:lncs26},
}

@proceedings{clef:ceur26,
  editor    = {Eva S{\'a}nchez Salido and Alberto Barr{\'o}n-Cede{\~n}o and Alba Garc{\'i}a Seco de Herrera and Sean MacAvaney and Julia Maria Stru{\ss}},
  title     = {Working Notes of {CLEF} 2026: Conference and Labs of the Evaluation Forum},
  booktitle = {Working Notes of {CLEF} 2026: Conference and Labs of the Evaluation Forum},
  series    = {{CEUR} Workshop Proceedings},
  publisher = {CEUR-WS.org},
  year      = {2026},
}

@proceedings{clef:lncs26,
  editor    = {Matthias Hagen and Martin Potthast and Benno Stein and Philipp Schaer and Eva Zangerle and Sean MacAvaney and Julia Maria Stru{\ss} and Eva S{\'a}nchez Salido and Alberto Barr{\'o}n-Cede{\~n}o and Alba Garc{\'i}a Seco de Herrera},
  title     = {Experimental IR Meets Multilinguality, Multimodality, and Interaction. Proceedings of the Seventeenth International Conference of the CLEF Association (CLEF 2026)},
  booktitle = {Experimental IR Meets Multilinguality, Multimodality, and Interaction. Proceedings of the Seventeenth International Conference of the CLEF Association (CLEF 2026)},
  series    = {Lecture Notes in Computer Science},
  publisher = {Springer},
  year      = {2026},
}

@misc{gemma4_2026,
  title        = {Gemma 4 Model Card},
  author       = {{Google DeepMind}},
  year         = {2026},
  howpublished = {\url{https://ai.google.dev/gemma/docs/core/model_card_4}},
  note         = {Accessed: 2026-06-03}
}

@misc{avatar2005,
  author    = {DiMartino, Michael Dante and Konietzko, Bryan},
  title     = {Avatar: The Last Airbender},
  year      = {2006},
  publisher = {Nickelodeon Animation Studios},
  note      = {Season 2, Episode 9: ``Bitter Work''}
}

\appendix
\section{Prompts}
\label{app:prompts}

\lstset{
  basicstyle=\ttfamily\footnotesize,
  breaklines=true,
  columns=fullflexible,
  frame=single,
  captionpos=b,
  xleftmargin=2pt, xrightmargin=2pt,
}

\subsection{Rationale Generation}

\begin{lstlisting}[caption={Generic rationale prompt. \texttt{\{query\}} defaults to ``General Wordplay''; \texttt{\{joke\_status\}} is ``IS a relevant pun/joke'' or ``is NOT a relevant pun/joke''.}]
Analyze the following text based on the search query "{query}".
Write EXACTLY ONE sentence explaining WHY it {joke_status}.
- If it is a joke, identify the specific linguistic mechanism (e.g., the exact pun, double meaning, or misdirection used).
- If it is not a joke, explain why it is merely a literal or unrelated statement.

Text: "{text}"
One-Sentence Rationale:
\end{lstlisting}

\begin{lstlisting}[caption={Typed rationale prompt. \texttt{\{query\}} is the actual query text retrieved from the training qrels.}]
Analyze the following text in the context of the search query "{query}".
Write EXACTLY ONE concise sentence explaining WHY this text {joke_status}.

If it IS a joke/pun, identify the SPECIFIC linguistic mechanism:
- Homophonic pun (words that sound alike but differ in meaning)
- Homographic pun (same spelling, different meanings)
- Compound pun (multiple puns in one text)
- Tom Swifty (adverb that creates a pun with the dialogue)
- Double entendre / double meaning
- Malapropism / word substitution
- Ironic twist / misdirection

If it is NOT a joke, explain: is it factual? definitions? unrelated topic? lacks wordplay?

Text: "{text}"
One-Sentence Rationale:
\end{lstlisting}

\subsection{Hard-Negative Generation}
The generic pipeline uses the first three prompts; the typed pipeline adds the fourth (near-miss pun).

\begin{lstlisting}[caption={Literal rewrite (Type 1). Rejected if character-level similarity to the original exceeds 0.85.}]
Rewrite this joke/wordplay as a completely literal, factual, non-humorous statement.
Keep the same core topic and key subject words. Remove ALL humor, puns, and wordplay.
Output ONLY the rewritten text (one sentence), nothing else.

Original: "{text}"
Literal version:
\end{lstlisting}

\begin{lstlisting}[caption={Defused joke (Type 2). Rejected if similarity to the original exceeds 0.95.}]
Take this joke and slightly change it so the punchline no longer works.
Keep the setup and structure, but replace the KEY word/phrase that creates the humor with a literal alternative.
The result should look like it COULD be a joke but isn't actually funny.
Output ONLY the modified text, nothing else.

Original joke: "{text}"
Defused version:
\end{lstlisting}

\begin{lstlisting}[caption={Wrong-topic joke (Type 3).}]
Write a short, original one-liner joke or pun about a COMPLETELY DIFFERENT topic than "{query}".
The joke should be genuinely funny but have NOTHING to do with the original query.
Output ONLY the joke (one sentence), nothing else.

Original topic: "{query}"
Unrelated joke:
\end{lstlisting}

\begin{lstlisting}[caption={Near-miss pun (Type 4, typed pipeline only).}]
Create a sentence that TRIES to be a pun related to "{query}" but FAILS.
The sentence should use a word that sounds SIMILAR to a pun-worthy word but isn't actually a pun.
It should feel like a bad/forced attempt at humor that doesn't land.
Output ONLY the sentence, nothing else.

Topic: "{query}"
Example of a GOOD pun on this topic: "{text}"
Failed pun attempt:
\end{lstlisting}

\subsection{LLM Judge}

\begin{lstlisting}[caption={Judge system prompt.}]
You are a humor and wordplay detection judge. You evaluate whether a text is relevant to a query AND contains humor, jokes, puns, wordplay, or any form of linguistic wit (double meanings, homophones, malapropisms, ironic twists). Answer only YES or NO.
\end{lstlisting}

\begin{lstlisting}[caption={Judge user prompt at inference. The YES/NO token logits are read directly to form the soft score $p_{\mathrm{YES}}$.}]
Query: "{query}"
Text: "{text}"
Is this a relevant joke? Answer YES or NO.
\end{lstlisting}

\end{document}